# Foundations for the Future:

## Institution building for the purpose of Artificial Intelligence governance

### Charlotte STIX[1]

**Abstract:** Governance efforts for artificial intelligence (AI) are taking on increasingly more concrete forms, drawing on a variety of approaches and instruments from hard regulation to standardisation efforts, aimed at mitigating challenges from high-risk AI systems. To implement these and other efforts, new institutions will need to be established on a national and international level. This paper sketches a blueprint of such institutions, and conducts in-depth investigations of three key components of any future AI governance institutions, exploring benefits and associated drawbacks: (1) "purpose", relating to the institution's overall goals and scope of work or mandate; (2) "geography", relating to questions of participation and the reach of jurisdiction; and (3) "capacity", the infrastructural and human make-up of the institution. Subsequently, the paper highlights noteworthy aspects of various institutional roles specifically around questions of institutional purpose, and frames what these could look like in practice, by placing these debates in a European context and proposing different iterations of a European AI Agency. Finally, conclusions and future research directions are proposed.

**Final accepted version:** https://link.springer.com/article/10.1007/s43681-021-00093-w

---

[1] Charlotte Stix, PhD Candidate, Philosophy and Ethics Group, Department of Industrial Engineering and Innovation Sciences, Eindhoven University of Technology, The Netherlands. Correspondence should be addressed to c.stix@tue.nl. ORCID iD: 0000-0001-5562-9234.

# 1. Introduction

The increasing capabilities of artificial intelligence (AI) systems,[2] and their manifold applications throughout society, have given rise to a range of governance concerns. New policies are needed to ensure the safe and reliable use and robust behaviour of these systems (Müller 2020; Calo 2017; Ulnicane et al. 2020), in accordance with fundamental and human rights (Raso et al. 2018) and ethical principles (Schiff et al. 2020). Recently, governments around the world have begun to approach the governance of AI through multiple levers (Cyman, Gromova, and Juchnevicius 2021). One timely example is the EU's horizontal regulation published in April 2021 which puts forward a regulatory framework for high-risk AI systems that are brought into or put on the Single Market (European Commission 2021). Another timely example is the Trade and Tech Council[3] established in the summer of 2021 between the United States and the European Union. One of the key policy areas for this Tech and Trade Council is to cooperate on the development of suitable standards for AI. Policy levers such as regulation or standardisation will play a crucial role when it comes to shaping the design, development and technical benchmarking of future AI systems on an international scale (Cihon 2019; Bradford 2020). The field of AI governance is relatively nascent and, as a result, there exist few dedicated specialist governmental institutions exclusively dedicated to supporting many of these initiatives. In order to properly develop, support and implement new AI governance efforts, such as legislative frameworks for AI, it is likely that a number of new institutions will need to be established over the coming years (Turner 2018; Almeida et al. 2021). It is therefore worthwhile to step back and think cautiously about the junctures we are about to lay (Pierson 2000) for the field of AI governance when developing these new institutions.

There are broadly two types of institutions that one could investigate: those that exist and may be adapted (i.e. either they organically evolve over time, *or* are 'repurposed' see (Alter and Raustiala 2018; Kunz and Ó hÉigeartaigh 2019; Stahl et al. 2021), and those that do not exist yet (but will eventually come into existence to fill the void that new governance measures have created and support them). This paper focuses on the latter and with a particular eye to those institutions set up by governments. This does not mean to make a judgement as to the relative importance of the work of NGOs in shaping AI governance, rather it serves to narrow the focus of this paper's investigation.

---

[2] This paper follows the definition of AI put forward by the European Commission's High Level Expert Group on AI (AI HLEG, 2019).
[3] See: https://ec.europa.eu/commission/presscorner/detail/en/IP_21_2990.

In doing so, it picks up from recent academic calls for an international governance coordinating committee for AI (Wallach and Marchant 2018), for an international regulatory agency for AI (Erdélyi and Goldsmith 2018) and for a G20 coordinating committee for AI governance (Jelinek, Wallach, and Kerimi 2020). Moreover, it takes up suggestions by scholars that at times it might be easier to add new institutions than to change or dismantle existing ones, if the latter cannot handle problems adequately (Alter and Raustiala 2018). Building on these calls and drawing on existing scholarship and insights, it addresses itself to those individuals who will be involved in setting up new institutions and those who are interested in conducting further research on pragmatic institution building for AI governance. The ambition of this paper is to empower them to make timely and suitable interventions in upcoming institution building efforts.

First, the paper outlines the rationale for deepening the investigation of institution-building for AI governance now, as well as the limitations to this paper's investigation. Subsequently, it puts forward three axes that contribute to the make-up of an institution: (i) *purpose*: what an institution is meant to do; (ii) *geography*: who are the members and what is the scope of jurisdiction; and, (iii) *capacity*: what and who makes up the institution. It breaks the latter down into infrastructural and human elements. Furthermore, under each subsection in (i), it highlights the pros and cons of various institutional roles and briefly frames what they could look like in practice. It does so by placing these roles in a European context and sketches a potential European AI Agency with them. Finally, conclusions and future research directions are put forward. Overall, the paper highlights pragmatic governance considerations and refrains from making normative assessments of these various institutional setups.

## 2. Motivation, Urgency and Limitations

Novel AI specific governance institutions working on soft governance mechanisms, that is non-binding rules, have already come into existence. For example, several intergovernmental fora such as the G7 or the OECD engage on mechanisms such as shared ethical principles (OECD AI Principles). One of the European Commission's foreign policy mechanisms, the  International Alliance for a Human Centric Approach to AI[4] which cooperates with like-minded countries such as Singapore, Japan or Canada to build on the European vision of a 'trustworthy AI' (AI HLEG, 2019). The Global Partnership on AI, through its 15 founding members and various dedicated working groups, aims to strengthen shared norms for AI through coordinated efforts. The OECD's ONE-AI working groups, adjacent to the OECD AI Policy observatory, supports the OECD recommendations to its members, ranging from policy measures to implementing trustworthy AI and compute. Prima facie these new fora appear well placed to support and engage in soft governance efforts.

At the same time there is mounting pressure to develop and implement stronger and more binding AI governance mechanisms than those covered by ethical principles, shared norms, and multi-stakeholder proposals. This pressure comes in a number of forms, such as from a societal viewpoint, accounting for the range of issue areas AI can cause (Butcher and Beridze 2019; Whittlestone, Arulkumaran, and Crosby 2021) and a technological viewpoint, accounting for the increasing capacity of AI and the speed of its development (Grace et al. 2018).  In response, for example at the trans-national level, the Council of Europe is currently examining what a legal framework for AI (international law) built on the Council of Europe's values could look like through its CAHAI committee (Ad Hoc Committee on Artificial Intelligence)[5]. Overall, as countries move towards harder governance efforts, such as regulation (European Commission 2021), standardisation[6] or certification they are likely to require increasingly specialised institutions.

---

[4] See:
https://digital-strategy.ec.europa.eu/en/funding/international-alliance-human-centric-approach-artificial-intellig
ence#:~:text=International%20alliance%20for%20a%20human%2Dcentric%20approach%20to%20Artificial%20I
ntelligence,-Opening%3A%2006%20November&text=human%2Dcentric%20AI.,common%20principles%20and%
20operational%20conclusions.
[5] See: https://www.coe.int/en/web/artificial-intelligence/cahai.

Moreover, as AI governance efforts soar and more coordination, action and policy proposals become necessary within a nation as well as at an international level, it is likely that there will be a need for more (and more specialised) governmental agencies to handle an increasingly diverse portfolio on top of existing files. Especially in cases where specialisation on a particular AI governance angle or AI application in a particular sector is required, and if that area is set to increase in volume, it is probable that the setting up of a specialised institution might be beneficial. It might be overall quicker, cheaper and more effective to build a new institution from scratch that is 'fit for purpose' rather than exert time, effort and political goodwill to change the structure of an existing institution.

Far from being a hypothetical proposition, this matters now: the EU is laying out extensive plans to set up new institutions for AI governance in the coming years (AI Act, 2021), NATO is putting forward plans for a civil-military Defence Innovation Accelerator for the North Atlantic, and many others will follow suit.

In April 2021, the European Commission put forward their horizontal risk-based regulation for AI, the AI Act (European Commission, 2021). In that AI Act, they described what form a regulatory framework for AI would look like in the EU and indicated a number of institutions that would partake in the fulfilment process of the regulatory requirements. While some of these already exist and will likely have their remit extended, such as market surveillance authorities, others, such as new institutions responsible for testing, certifying and inspecting AI systems (notified conformity bodies) are likely to be established to cope with an increasing demand once the AI Act is enforced. Moreover, while the AI Act does not foresee one single supervisory Agency for AI in the EU, other EU institutions like the European Parliament have advocated for such a new Agency. The 2017 resolution on 'Civil Law Rules on Robotics' (European Parliament, 2017) and the 2019 report on 'A comprehensive European industrial policy on artificial intelligence and robotics' (European Parliament, 2019) both called for the establishment of a European AI Agency that would be in charge of supplying technical, legal and ethical expertise and intervention. Given the sheer scope the new horizontal legislation for AI will introduce and the number of associated institutions, networks and processes that need to be either set up or managed, and given this political direction from the European Parliament, it is not unlikely that a European AI Agency will be on the horizon within the coming 10 years. This institution is likely enough to be established, yet far enough in the future for there to be meaningful interventions to its structure. It will be used as a practical example under each axis in Section 3.i to demonstrate how various considerations could play out.

---

[6] See for example work on AI standards by the ISO/IEC JTC 1/SC 42 Committee. Available here: https://www.iso.org/committee/6794475.html.

Different institutional set-ups will yield different path dependencies. Those that do get put into motion over the coming years are likely to be especially critical (Stix and Maas 2021) because they form the lens through which governments will be able to interact with progressing AI technologies and enact appropriate AI governance measures. This wouldn't merit outsized concern, if it were possible to clearly forecast AI progress across various sectors and if there was common expert consensus about the development path of AI over the coming decades. However, even experts struggle to agree (Grace et al. 2018; Müller and Bostrom 2016) and given the ubiquitous nature of AI it is difficult, if not impossible, to forecast now what AI governance frameworks might be needed in the near or far future. In addition, existing institutions, their policy texts and legal frameworks are historically often drawn from in moments of unexpected change (Stix and Maas 2021), because it can be quicker to react with what exists than develop something new.

Once new institutions are established they are difficult to change (Sanders 2008). Equally, it is difficult to adapt larger international institutions to changing issues of concern within the landscape (Morin et al. 2019) or to shift from the processes that were originally conceived to different mechanisms and fluctuating missions (Baccaro and Mele 2012). This all goes to say that, where new institutions may be the vehicle of choice to action current policy measures, in the future, the composition of these new institutions will itself precede and, crucially, *constrain* or *catalyse* future governance approaches. Early stage decisions to establish new institutions, or the choice to forego such new institutions (Cihon, Maas, and Kemp 2020), are all likely to have a downstream, or lock-in, effect on the efficiency of government measures and on the field as a whole. It is therefore timely and important to think about institution building as one of the critical path dependencies we are developing now.

Prior to delving into a deeper investigation of institutional set-ups and associated considerations, we place the work within the existing literature and outline the scope of the work set out in the remainder of the article. The paper builds on and contributes to existing literature on the topic of institution building (Goodin 1998; Koremenos, Lipson, and Snidal 2001a; Koremenos, Lipson, and Snidal 2001b) by focussing on the topic of institution building and design for AI governance interventions specifically. In particular, the paper adopts the lens of historical institutionalism (Thelen 1999), suggesting that for AI governance new institutions will be built dependent on the current promise and peril of (and the associated mechanisms to enable or minimise) AI. Newly established AI governance institutions will themselves shape political interactions as those interactions will be a result of the newly established institutions and their processes in turn (Sanders 2008). The question of how to set up AI governance institutions therefore merits further scholarly investigation.

Accordingly, section 3 puts forward a selection of axes that need to be considered in building new AI governance institutions: (i) *purpose*, (ii) *geography*, and (iii) *capacity*. These provide a framework for envisioning different types of institutional designs and their relative merits and shortcomings.

This preliminary set of axes does not claim to be exhaustive. Neither can each individual investigation fully account for the range of interactions, interdependencies and considerations that may (or should) exist. Inevitably, the creation of institutions for AI governance will have a political and geopolitical dimension. Angles such as those of cooperation theory are relevant to that dimension, but fall outside of the scope of this paper.

# 3. The main axes: purpose, geography and capacity

The following investigation maps the axes of *purpose*, *geography* and *capacity*, with particular focus paid to *purpose*. It serves to outline a first framework towards institution building for AI governance. How the aspects are combined, what is favoured and what isn't, will depend on the values, difficulties and models of the future/AI governance worldview each reader holds, as well as on the political, societal and economic climate those setting out to establish a new institution will find themselves in.

## 3.1. Purpose: What is it meant to do?

Among the first things that will need to be decided upon is the *purpose* of the new institution. That is: *what is it meant to do*? Prima facie, this might seem like a straightforward question, yet the following paragraphs will evidence why neither the question nor the answer(s) are. Each subsection introduces the outline of a role an institution for AI governance could take: *coordinator*, *analyser*, *developer* and *investigator*. Each subsection introduces the relevant role, followed by examples of existing institutions that match this model (where appropriate), a discussion of some relevant considerations, and concludes with transferring the explored role to the case of a potential European AI Agency, demonstrating what shape this might take in practice.

Finally, this section concludes with a short account of hybrid cases between the various roles, briefly accounting for how they might intersect or support each other in practice.

## 3.1.i. The coordinator institution

Current proposals by academic researchers (Wallach and Marchant 2018; Jelinek, Wallach, and Kerimi 2020) and governments (such as the European AI Board, European Commission, 2021) often suggest something akin to what this paper terms a *coordinator*: an institution whose purpose it is to coordinate between a number of actions, policy efforts or norms.

*What does it do?*

Coordinator institutions could, for example, work with the rising number of ethical guidelines (Zeng, Lu, and Huangfu 2018) and attempt to operationalise them more clearly. They could also serve as an umbrella organisation to coordinate activities across like-minded nations (Erdélyi and Goldsmith 2018), helping different groups to learn lessons from one another and avoid duplicating efforts.

Moreover, there is a rising number of relevant but often uncoordinated efforts to tackle certain aspects relevant to AI governance, such as certification schemes (Winter et al. 2021), testing procedures (Brundage et al. 2020), or setting out shared definitions of 'meaningful human control' across diverse contexts. These too could benefit from coordination to increase efficiency, coherence and policy impact. A coordinator institution (others have referred to similar roles as 'orchestrator' cf. (Abbott and Snidal 2010; Abbott and Genschel 2000), would focus on encouraging the exchange and synchronisation between institutions or efforts, amplifying and streamlining work done elsewhere. It could fill a current gap 'in the market' of some aspects of AI governance, stemming issues that arise from quickly developing and independent streams of efforts, and combat a fragmented landscape as well as norm conflict (Garcia 2020).

*What models of coordinator organisations exist?*

Intergovernmental organisations are the closest example of a coordinator role for the purpose of AI governance. While they are much more complex and encompass some roles discussed later in this section, intergovernmental organisations such as the United Nations and organisations covered thereunder such as the ITU; or the G20 or the NATO can be seen as coordinators. They predominantly act to coordinate efforts, strategies and common interests under international law and are composed of sovereign states (actors) that share mutual interests or seek a neutral platform for discussion and exchange. From a European perspective, the European Data Protection Board coordinating between various national data protection agencies could be seen as a coordinator.

*What are relevant considerations?*

*Power*

One of the benefits of a coordinator could be that it serves as a relatively neutral environment within which various members (e.g. groups, nations) are able to discuss and develop broad agreement on shared initiatives. It may also serve to alleviate the workload of its members by acting as a 'quasi back office' supporting synchronisation, distribution and organisation of relevant work streams.

The coordinator's main purpose is to serve to coordinate for its members and not to hold an independent political agenda. Nevertheless, it should not be mistaken as being completely powerless or unable to (inadvertently) shape the prominence of certain governance efforts.

Depending on how it is set up a coordinator is either *independent* to its members, or the coordinator *is* the members. Similarly, depending on the procedures established through which it can act, either version would have decision making power over the approval of new members and the decision making power over new coordination efforts within its remit. In either case, whether willingly or unwillingly, the coordinator will inherently only amplify the mutual interest of its members vis-a-vis the international political stage.

This can lead to exclusion and oversight, uninformed or ill-informed choices to develop or deploy AI systems which may negatively impact nations or groups that do not form part of the coordinator (Hagerty and Rubinov 2019), or an amplification of approaches that might be suited for some, but unsuited for the entirety of the international community (Schiff et al. 2020).

Moreover, the act of coordination itself comes with trade-offs: navigating a common line among efforts or approaches might result in some falling outside of the plotted line. This would translate into a decrease of support from the coordinator for those efforts given they do not match the majority line. On the other side of the coin, those efforts or approaches with more resources to start with are likely to have a higher ability of dominating the common line, or even to hedge their bets and participate in parallel in more than one coordination regime.

*Access*

Questions regarding membership procedure should be addressed while setting up the institution or be clearly related to the topic of activities the coordinator is established for. For example, when the EU discusses coordination and cooperation with 'like-minded countries' through its external policy vehicle, the International Alliance for Human Centric AI, then this narrows the scope of engagement to those countries that follow a similar adherence to fundamental and human rights, and values. While access to the coordinator should be possible to all relevant actors in order to ensure diversity and representation in value and opinion, quasi open access does come with a trade off. For a coordinator, more members might result in more time spent on early stage coordination interventions and mutual understandings, before time can be spent on actioning those understandings and streamlining efforts.

The coordinators' ability to fulfil their ultimate duty  and to add value as an institution becomes increasingly difficult the thinner the coordinator is spread across increasing efforts, novel angels and shifting needs. This could also contribute to slower reactions to and adaptations to potentially quickly shifting technical landscapes (Marchant, Allenby, and Herkert 2011).

*Focus*

The focus of the coordinator, and its actions should be timely and appropriate. The focus will likely be dictated by the initial group of members. Timely and appropriate focus also encompassess a certain degree of agility. This might be especially important to bear in mind if and when the coordinator expands, and in light of foreseen and unforeseen shifts and needs within the AI governance landscape. For example, in response to an unexpected crisis such as the COVID-19 pandemic previously known issues such as those of privacy and security in light of technological tools might become imminently pressing and require fast coordinated responses (Tzachor et al. 2020) that cross national boundaries.

Depending on the flexibility the coordinator is endowed with, either independently from its members or through e.g. voting procedures from its members, the initial set up could have an outsized impact on the future agility and responsiveness of the coordinator. It could provide an 'first mover advantage' to those that set up the institution and set its first directions.

*How could this look like in practice?*

A future AI Agency in the EU might take the role of a coordinator. In practice, this could mean that the 27 EU member states would be the founding members of that agency, as they are the only actors that are directly involved with the EU's governance efforts such as regulation. In that sense, it would be sufficiently inclusionary and is unlikely to need to structure elaborate access procedures in the near future (until a new country joins the EU). While its scope may not be stretched through access of new members, it might nevertheless stretch itself thin quickly on actions, unless there is a narrowly defined scope for the coordinator EU AI Agency. The large number of members might also lengthen reaction times under unforeseen circumstances.[7]

Currently, a multitude of efforts are coordinated between member states through the European Commission and under the helmet of the Coordinated Plan on AI (European Commission 2018; European Commission 2021). In addition, new challenges that a coordinator could tackle will arise out of the AI Act (European Commission 2021) and its implementation across the EU. The coordinator EU AI Agency could pick up coordination efforts under the Coordinated Plan on AI (European Commission 2018; European Commission 2021) such as aligning AI strategies, pooling resources and strengthening the ecosystem between member states. Or, it could act as a coordinator EU AI Agency for the AI Act supporting the coherent implementation, enactment and functioning of the horizontal regulation for AI.

---

[7] See Commission President von der Leyen's statement describing the EU as a tanker, whereas an independent country such as the UK can act as a speedboat: https://www.theguardian.com/society/2021/feb/05/ursula-von-der-leyen-uk-covid-vaccine-speedboat-eu-tanker.

Given aforementioned considerations, the highest benefit might be derived if its focus is clear cut and does not overlap with existing efforts such as those already led by the European Commission. As such, coordination on the AI Act (European Commission 2021) and either (i) coordination between (a) all relevant institutions within the member states' landscape (e.g. national supervisory authority, notifying authorities or notified bodies), (b) between a subset of similar institutions within the member states' landscape such as notified conformity assessment bodies; or, (ii) on specific topics such as X, could derive the highest benefit.

It would mean that there is one coordinator that is tasked with monitoring, implementing and supervising relevant activities with regards to the AI Act and/or a number of smaller corresponding bodies across the EU. This could strengthen information exchange and increase the speed at which scope specific problems can be identified and addressed, ultimately supporting the higher order goal of a well functioning regulatory framework for AI.

## 3.1.ii. The analyser institution

Another role for a future AI governance institution could be that of an analyzer.

*What does it do?*

An analyser could fulfil several roles: it could serve to map existing efforts and identify gaps across various governments. The European Commission, for example, undertook such mapping efforts in the Coordinated Plan on AI: 2021 review (European Commission 2021) to identify where specific AI-relevant measures had not yet been implemented or need to be established across the EU. It could compile data sets and information on the technical landscape and sketch technological trajectories. The AI Index[8], for example, tracks and publishes data corresponding to progress within various AI applications and research areas. It could collate relevant information about the opportunities and risks associated with the use of AI within specific sectors along the lines of the Centre for Data Ethics and Innovation's (CDEI) AI Barometer[9] work. In short, an analyser draws new conclusions from qualitative and quantitative information.

*What models of an analyser exist?*

In addition to the aforementioned examples, where in the case of the European Commission and the CDEI the analyser role of those institutions is housed within a larger institutional framework holding multiple roles, a good example of an analyser institution is the European Parliament's Think Tank.[10] It provides studies, briefing and in-depth analyses on a variety of topic areas to the members of the European Parliament and makes them publicly available. These can be requested from the European Parliament or developed subsequent to a particularly pertinent happening. In doing so, the European Parliament Think Tank supports the well-functioning of the European Parliament, the speed at which relevant decisions can be made and heightens the understanding of participants on relevant key topics. Another example would be the European Commission's Joint Research Centre's AI Watch[11] which monitors, and provides high-level analyses about, research, industrial capacity and policy initiatives across the EU to inform the European Commission on policy decisions.

*What are relevant considerations?*

---

[8] See: https://hai.stanford.edu/research/ai-index-2021.
[9] See: https://www.gov.uk/government/publications/cdei-ai-barometer.
[10] See: https://www.europarl.europa.eu/thinktank/en/home.html.
[11] See: https://knowledge4policy.ec.europa.eu/ai-watch/about_en.

The role of an analyser is more active than that of a coordinator, in that it interferes more directly with the governance or policy making process by way of providing crucial information that can inform and shape those decision making processes.

*What or who does it respond to?*

One key consideration is whether an analyzer is established in response to an identified need within the governance landscape or whether it is established without a clear angle in mind but as an independent institution to provide services to advise decision-makers. In the first case, the analyzer would be directly driven by its scope and gain its direction from the event and those who chose to establish it as a response. The second case is a lot broader and such an institution could, in principle, allow for a more diverse range of ad hoc analyses, be that on specific sectors such as the automotive sector, the financial sector or the healthcare sector, or on specific topic areas such as compute or data. The European Parliament's Think Tank would fall into this category. Instead of serving to inform one particular direction, it adapts to the shifting needs of those who established it or those who it is supposed to inform, be that high-level individuals, governmental agencies or entire governments. It might therefore be able to cover more breadth. However, one trade-off might be in-depth subject expertise. An analyser focussed on one specific topic area could quickly become an expert institution in that field, whereas an analyser focussed on multiple topics might need to draw on outside expertise for particular areas which could cause additional time and effort.

*Independence*

Using the aforementioned examples, it would appear that an analyser can either be (i) dependent on those that demand its work, or (i) independent to those that demand its work. In both cases, though to a varying degree, the quality of the work of the analyser will depend on the range of relevant information it has access to. In the case of working with publicly available information this ought not to pose a problem, however, if it concerns information that is predominantly derived from non-public sources then the analyser is dependent on the information provider to ensure that it is complete, accurate and comprehensive. Where the information provider may overlap with the group that asks the analyser to conduct work on their behalf this can become tricky as the result of the analysis might be (inadvertently) shaped by the actor that requests the work. This could even hold true where no information is exchanged but where the framing of the work, or question, is given by the actor requesting the service and not developed by the analyser independently, (inadvertently) narrowing the scope of research and shaping its direction. Moreover, in these cases, existing narratives might be amplified instead of empowering the discovery of novel, equally pressing ones (Kak 2020).

*Shape of the final product*

Beyond the process which the analyser undergoes to develop its final piece of work, another consideration is the actual shape of that work.

That means, the final product could be as 'shallow' and 'inactive' as identifying shared issues and solutions, or more towards a 'deeper' and more 'active' product by providing coherent policies or sets of requirements to be implemented based on the work undertaken. The depth of analysis and the ease with which it can be actioned or implemented matters to the impact the final product of the analyzer will have. Whether these suggestions are binding to the actor(s) that requested the analyser to undertake the work, and the degree to which they need to act based on the information received will differ. On that spectrum, between non-binding suggestions with no need to act on them and either binding suggestions or those where action and change are a clear expectation and requirement, the analyzer's importance within the AI governance ecosystem will differ.

Put differently, the higher the likelihood of its work being adopted in decision making processes will be, the more influential it is and therefore this will likely affect previous questions surrounding its (political or financial) independence. In a situation where policy makers and governments rapidly look to develop new approaches towards the governance of AI, analysers hold some non-negligible power. For one, they can influence those looking towards finding solutions, measures and new information under the pressure of addressing rapid tech development in comparison to slow policy making filling a void in current policy making (Marchant, Allenby, and Herkert 2011).

*How could this look like in practice?*

For the hypothetical case of a European AI Agency, one version of an analyser role could be independent from those that request its work. It could serve as a third party compiling data on specific efforts which it can then offer to various institutions within the ecosystem in the form of recommendations in order to change or amend their actions when it identifies pain points. For example, in the case of the implementation of the horizontal regulation for AI (European Commission 2021) if the analyser identifies that assessments of some high-risk AI systems are slow across the EU for a specific subset of the third party conformity assessment procedure it can recommend to increase the number of notified conformity assessment bodies on that topic. Depending on how it is set up, its work could be both binding and non-binding which would influence its ability to meaningfully shape the pain points it identifies.

On the other hand, an analyser role for a European AI Agency could also be dependent on those that have set it up, responding to various different requests for its work from a range of actors, such as surveys on AI uptake, statistics on overall adherence to the regulatory framework or divergences between member states. In this case, the analyzer would likely respond to the European Commission and member states or a subset of the institutions across the EU that are involved in the regulatory framework, which might narrow its scope to discover novel pain points in accordance.

In each of these two cases, the motivation of the analyser's work will be different and either depend on the priorities it has identified for itself or on the priorities of the groups it represents.

Finally, a European AI Agency could also take the role of an AI observatory. One version could be an AI observatory that is set up by EU institutions and the governments of the member states but sufficiently independent to provide meaningful insight into the actual state of AI development and deployment across the EU in addition to existing work. Such an AI observatory could also double down on quantitative analysis only, to gain a competitive advantage over existing work in that space in the EU. It could track, measure and eventually forecast the progression of various AI technologies, map relevant aspects of their supply chains or components to develop AI systems (e.g. compute), and monitor capacities and impacts of deployed AI systems across the EU.

## 3.1.iii. The developer institution

Most versions of an analyser institution will stop short of coming up with clear proposals to implement and action its analysis.

*What does it do?*

Where an analyser might identify gaps in the technical landscape, a developer would structure policy proposals to close the gaps. Where an analyzer might map adherence to certain principles or ambitions, a developer would argue which ones are more important and give advice on how to strengthen those. Where an analyzer would suggest that there is a lack of sufficient institutions to account for a certain procedure needed within the ecosystem, e.g. for conformity assessment for the AI Act (European Commission, 2021), a developer would propose which institutions should fill the gap and how they can do so in a timely manner.

A developer provides either directly actionable and implementable measures and advice, or formulates new policy solutions to existing issues. Its political standing would be such that the solutions it puts forward have a high probability of being adopted by decision-makers, setting it aside from other institutions such as a variety of think-tanks.

The developer role brings an interesting power with it in that it can become a creator of novel policy measures, economic and financial decisions and legal proposals. The development process itself might include multiple steps, for example from the identification of a pressing policy issue to the proposed solution.

For example, the AI Act (European Commission 2021) was developed and proposed within one institution within the EU, the European Commission. The development of a regulatory framework was achieved within the same institution that explored risk matrices, case studies and that supported the development of principled guidelines (AI HLEG 2019), which all resulted in crucial information and suggestions included within the final proposal, the AI Act.

The type of institution proposed in this paper is envisioned more independent than this. While ensuring continuity and coherency throughout the development process of a given policy is important, it appears equally important to have some degree of political independence to ensure that the suggestions made are not only wanted but also needed.

*What models of developer organisations exist?*

Traditionally, the development of policy proposals or the development of actionable and implementable solutions  is shared between governmental agencies, departments and senior government officials such as in the UK context (Waller 2009), or in the EU context between the European Commission, other relevant EU institutions where applicable, and member states (Wallace et al. 2020).

As sketched in this paper, however, a developer would almost serve as an external circuit breaker to ensure that the work undertaken by governments is timely, comprehensive and sufficiently in depth. In that sense, it might take the role of examining blindspots and proposing solutions for those by way of its own initiative, in addition to work it might be asked to undertake by various government agencies. Of course, it is important that in either case the developer would be in a position such that its work will be actioned or meaningfully acted upon once proposed.

*What are relevant considerations?*

It is probable that it is difficult to set up a developer in a manner that makes it reasonably agile to account for what is needed in a rapidly changing policy space (Cihon, Maas, and Kemp 2020). For one, it appears that this would require a considerable amount of foresight and time spent on thinking through various future scenarios to ensure that the institution is set up in an appropriate and adaptable manner.

*Future orientedness*

In order to ensure flexibility and a sufficient degree of future orientedness, it might serve the developer to have a mandate beyond election cycles and independent to political agendas. This consideration orients itself on the case of Rights for Future Generations and the difficulty to account for future-oriented policy making in governmental structures that haven't been set up for that purpose. Over the past year, several governments have attempted to set up a Commissioner for Future Generations within governments, to oversee various policy processes and ensure that the voice of future generations is at the table when crucial decisions are made. While it proved not only difficult to effectively 'add on' this new role in government, most of these efforts failed after a short time (Jones, O'Brien, and Ryan 2018). It is timely to consider an institution's ability to act with an eye towards a longer time horizon independent to the political agenda of the day, as AI is likely to be increasingly influential across various political areas in the near and longer-term future (Raso et al. 2018; Sharkey 2019; Molnar 2019; Nemitz 2018) and will significantly disrupt many processes we have grown accustomed to as a society (Buolamwini and Gebru 2018; Anderson et al. 2014; Russell 2019).

*How could this look like in practice?*

One shape this could take for a hypothetical EU AI Agency is that it forms an independent institution with regard to the European Commission and those located within their respective Member States. While its scope could vary, an interesting option would be for it to work on the proposals for new high-risk AI systems to be added to the legislative framework. It could act as a foresight body to EU institutions, independently reviewing and analysing the landscape and, from that work, distilling noteworthy technological developments and proposing new high-risk AI systems in a timely manner. Its work could directly inform the adaptation procedure for new high-risk AI systems to be added to the legislative framework in collaboration with the European Commission, the European AI board and other relevant expert groups.

## 3.1.vi. The investigator institution

Finally, the last role of an institution this paper sketches is that of an investigator.

*What does it do?*

An investigator institution might track, monitor and investigate efforts or audit actors with regard to adherence to specific hard governance structures. It captures those abilities typically associated with a 'watchdog'. In short, it investigates whether or not actors such as governments, companies or specific organisations adhere to the relevant standards, procedures or laws. In doing so, it also serves as an external motivator for relevant actors to ensure that they are behaving ethically.

It should be noted that investigatory ability (and the time, resources and efforts expended on it) would likely be predominantly warranted or needed only when a measure has crossed the threshold between soft and hard AI governance.

*What models of investigator organisations exist?*

There are several 'watchdog' organisations at international level for example those monitoring human rights abuses such as Amnesty International[12] or the Human Rights Council.[13] A related model to the proposition of an investigator at European-level could be the European Court of Auditors[14] or the European Ombudsman[15] who both work to ensure that the EU is transparent in its workings and accountable for its actions. Citizens and organisations alike can file a complaint with the European Ombudsman against the EU's administration in cases of misconduct which will then be investigated. The European Court of auditors assesses how taxpayers money has been spent and reports this to the EU institutions and citizens.

---

[12] See: https://www.amnesty.org.uk/.
[13] See: www.ohchr.org.

A US-based example of what an investigator-type role could look like could be the independent and non-partisan Inspector Generals who conduct reviews within various Federal Agencies. They play a crucial role when it comes to government oversight and can conduct "audits, investigations, inspections, and evaluations"[16] into agency programs.

*What are relevant considerations?*

*Set up*

The set up of such an investigator will largely depend on the number of topics it is expected to cover. The options could range from one big investigator institution for the whole topic of AI itself with multiple sub teams for specific aspects, adding teams as needed, to multiple smaller institutions ready to investigate one specific element (assuming a break by skillset, sector, etc) under a coordinator. Similar considerations and trade-offs apply when putting the role in an international context assuming that many hard governance mechanisms such as regulation will apply to more than one country. The benefits and trade-offs of one big impartial investigator versus multiple nationally located investigators needs to be weighed carefully.

The setting up of an investigator includes political considerations and if it is expected to function as a third party to those it investigates, such as governments, then it should remain sufficiently independent and impartial to those it investigates. This could become a balancing act, if the investigator equally needs to ensure that it has access to all relevant and up to date information to correctly fulfil its tasks. This is where investigative power comes into play.

---

[14] See: https://www.eca.europa.eu/en/Pages/ecadefault.aspx.
[15] See: https://www.ombudsman.europa.eu/en/.
[16] See: https://fas.org/sgp/crs/misc/R45450.pdf.

*Investigative Power*

It makes a big difference to the impact of the investigator's output whether the investigator reviews information and data that is publicly available (or curated information made available upon request, relying on the goodwill of the institution that is being investigated), or whether it has the legal power to request access to all documentation, data, information, audits etc. including those kept internally. The investigative power itself can fall on a spectrum. The investigator may have a broad scope regarding one specific area, such as the correct implementation of legislative efforts, in which case specific investigative actions might need to be further justified before they can be undertaken. Or, it could be responsible for the investigation of a narrow scope, such as whether or not specific companies adhere to a legal criteria, in which case it is clearer what falls within its scope of investigatory power and what does not. Finally, it should be considered whether an investigator also holds the power to reprimand actors that fall afoul their obligations or whether that power should be vested in another institution.

*How could this look like in practice?*

One version of an EU AI Agency that acts as an investigator could be set up as a supranational agency encompassing all member states, yet as an independent actor to those member states. Depending on its size and capacity (technical and human) it could either act in response to requests made by individuals, groups or governments, to conduct ad hoc investigative exercises for a particular area, or have it within its remit to take independent action on a particular area or areas falling under its scope. For example, its scope might cover investigating the work of sub-contracted third party conformity assessment bodies that undertake assessments for AI systems that are parts of products entering the EU market, to ensure that they all have equally high standards. Given a pre-existing threat of fragmentation between various aligned efforts within the EU (Stix 2019), such a supranational AI Agency could support a coherent application of future legislative instruments and policies, while remaining politically independent.

## 3.1.v. Additional considerations

It is clear that, in the real world, the roles sketched in the previous paragraphs are unlikely to be completely independent of one another, and if they are, that would lead to a different set of problems (or benefits). Moreover, there are good surface-level arguments to be made about the benefit of mixing them, such as a centralisation and therefore streamlining of processes, shaving off potential losses that result out of time and effort spent on communicating, updating and navigating between independent organisations and an avoidance of polluting the landscape with an increasing number of individual institutions (Cihon, Maas, and Kemp 2020). On the other hand, it might be difficult for a larger institution to adapt to novel challenges in comparison to a network of independent institutions.

While the roles have been presented independently of one another, one could see the institutions sketched as forming links in a larger chain, increasing in complexity, power and impact. In the aforementioned cases several of the institutions could intersect with one another and take the form of hybrid institutions. For example, an analyser that becomes specialised in a particular topic area could benefit from either morphing into an analyser/developer hybrid or from being closely associated to a developer. Similarly, a developer/investigator hybrid may have the benefit that it has an exceptional understanding of what adherence to a specific measure would look like or not, given that it developed it, whereas independent investigator institutions may need to continuously source expertise from an independent developer institution.

Finally, it should be mentioned that there are a variety of further areas of competency that an institution can be built for that merit investigation, such as those of an enforcer, for example something such as the European Court of Justice. Unfortunately, this is outside the scope of this paper.

The paper now moves towards an exploration of the *geographical* considerations and *capacity*, complementing the landscape sketched this far.

## 3.2. Geography: Who are the members and what is the scope of jurisdiction?

It is increasingly evident that AI-systems will impact society well beyond their original place of development or deployment (Brundage et al. 2018). Simply put, they do not respect national borders. Therefore, many AI governance concerns could be seen as multi-country concerns.

Some of the broad considerations for this axis are: What is the benefit or downside of a new multi-country institution? How does this fare in comparison to nationally 'restricted' institutions? Another consideration with regards to geography will depend on the type of AI-systems that are to be governed: for example, instances where cross-border infrastructure is desirable (e.g. autonomous vehicles) will need a different approach than those where AI-systems are deployed in public services of individual nations, that is reasonably 'restricted' to one country. In the former case, a multi-country institution might enable coherent testing procedures, common protocols, legislative efforts and smooth operation of AI-systems between affected nations. The remainder of this section will contribute pragmatic perspectives to existing literature on the topics and concerns (Cihon, Maas, and Kemp 2020; Wallach and Marchant 2018).

*Access, Inclusion and Participation*

Decisions made under the geographical axes will either reinforce and mirror, or reshape existing political alliances, "like-minded partnerships"[17] or governance efforts. A multi-country institution must therefore consider questions of access, inclusion and participation. As Koremenos et al. (Koremenos, Lipson, and Snidal 2001b) outline: is access inclusive by design, restricted to specific states that share certain commonalities, regional or universal? Moreover, how should the institution (be able to) handle a shifting geopolitical landscape *or* expansion (including inside or outside pressure for expansion) ?

In any case, an institution with multiple countries will evangelise the chosen direction by those countries on an international level by virtue of amplifying the countries' vision, disseminating it and pooling resources to expand on it. This might lead to a competitive advantage for those nations within the institutions, who often are already comparably more powerful, effectively steamrolling efforts in individual nations that are not part of this (larger) group (cf. this phenomenon in AI Ethics Principles (ÓhÉigeartaigh et al. 2020; Hagendorff 2020; Mohamed, Png, and Isaac 2020). For example, a new institution that holds any of the roles sketched under Section 3 which is predominantly populated by western actors could inadvertently outmanoeuvre concerns and efforts in other global regions by way of heightened visibility and amplification of the former's voice. This makes such an institution a partial actor on the global playing field and questions of access (and power) pressing.

Moveover, this scenario might also 'tip the balance': it could 'force' nations to join the effort despite it not being in their best interest given their particular ecosystem, where their alternative would be the role of a complete outsider to the seemingly new global network that is getting built. Finally, decisions to expand the membership to an institution (or not) could turn into political provocations (on purpose, or not).

At the same time, if several nations (in broad agreement) expect that their position towards

---

AI governance is broadly more beneficial than that of other nations, it may be a reasonable political and future-oriented decision (from their perspective) to cooperate, coordinate and pool efforts between them through a dedicated institution. In short, if your belief was such that there is a threat to good AI governance from several increasingly powerful nations, one of your approaches may be to pool together with those nations that align with your vision of AI governance in order to not only level the playing field but to gain a competitive advantage and use it to the benefit of all. Whether or not that belief is correct or whether one way is more beneficial than that of another country (and which) is outside the scope of this paper. This might also contribute to a reinforced spill-over effect of the chosen and predominant path where more nations sign up to signal that they too wish to demonstrate adherence to certain (broadly beneficial) governance mechanisms.

Conversely, if nations or bigger institutions chose not to form a new institution, a proliferation of similar but distinct institutions could affect fragmentation of global AI governance regimes (and associated governance regimes). Distinct 'regimes' which have little mutual coordination, cooperation or shared expertise, might overlap or even clash. Beyond that, nation specific institutions may hinder, or complicate, a corresponding 'scaling' of AI governance measures in light of increasing AI capabilities in the coming years and decades (Müller and Bostrom 2016). To match this, and to be able to comprehensively navigate AI governance and make new institutions future-proof regardless of their geographical make-up it is likely that technical capabilities and expertise will play a crucial role. To that end, the next section will briefly investigate the third axis of this paper: *capacity*.

## 3.3. Capacity: *What* and *who* forms part?

This subsection introduces infrastructural considerations with regard to AI governance institutions. It is divided into *technical* and *human capacity*, though both stand in close relation. *Capacity* relates to the previous two axes (*purpose* and *geography)* in that the *what* and *where* of a given institution, will influence what the institution needs in terms of capacities for it to thrive, both from a technical and non-technical perspective.

This paper proposes that access to technical infrastructure could play an important role for future AI governance institutions. Governance proposals, policy suggestions and requirements could be improved and tailored to the state of the art (AI Index), possibly combatting the pacing gap (Marchant, Allenby, and Herkert 2011), if the institution has capacity to accordingly run their own tests, measurements and map AI progress (see e.g. work by the AI Watch). This could range from access to compute (Brundage et al. 2020), increasing available data sets (European Commission 2021), to testing and experimentation facilities (European Commission 2021). It can minimize bottlenecks in terms of information exchange, knowledge of opportunities and risks, and timeliness and increase speed between what is to be governed and the associated governance actions, decisions and proposals themselves. This could contribute to more agility, specificity and foresight in policy making for AI.

Moreover, access to technical infrastructure in-house can enable those who are developing proposals, investigating measures or otherwise, to 'fact check' ideas, possibilities and limitations without excessive reliance on third parties to provide this information to regulators or other key decision makers. Indeed, a total lack of access to technological infrastructure in house to e.g. verify claims about technical possibilities, can hand-off significant power or influence over governance decisions to other actors. These can end up becoming the sole source of information for policy makers as to what is and isn't possible to do with the technology, and are therefore capable of shaping governance measures (possibly to their interests). The power balance might not be levelled (and it might not need to be or desirable to completely do so)[18] with some degree of in-house ability to test, develop and trial run procedures but it could be significantly readjusted to benefit the suitable development of AI governance from the perspective of governments. Such a readjustment could even benefit technology companies as it is likely that hard governance efforts will have less of a 'knee jerk' quality and be more nuanced, implementable and timely than otherwise.

---

[18] This paper does not mean to suggest that governmental institutions should become tech companies. It merely suggests that some in-house technical capacity (and direct expertise) can be useful for the purpose of good AI governance.

Infrastructure of all forms, including technical, needs individuals who can understand, handle and extract meaningful information from it: the human capacity.

For future oriented and informed work on AI governance, it is vital to be able to reasonably comprehend, foresee, evaluate and measure a variety of scenarios with regards to the technology. This needs both technical and human capacity. Building up human capacity could take broadly two forms: (a) out of house capacity with either (i) a network of individual experts to draw on when needed[19], or (ii) an expert groups and external panels[20] and (b) in house capacity where you build up a sufficiently sized team with a range of relevant expertises such as technical, legal, and ethical, as well as diverse backgrounds, in the first place. Both (a) and (b) need never be considered mutually exclusive. For example, when soliciting expertise from external groups or networks, how can it be ensured that staff would be able to ask pertinent questions in the first instance? The answers to this question will correlate with (b) hiring diverse expertises and backgrounds: it will depend on the expertise with which the institution is populated, the degree to which individuals have the opportunity to keep updating their expertise, learning relevant new information, and engaging with technological progress.

Diversity for the purpose of this paper accounts for two things: a diverse range of expertise (including technical, legal, STS backgrounds and more), and a diverse range of backgrounds (including socioeconomic, ethnic, political and more). This is needed to cross-pollination ideas, solutions and recommendations that work for all within society, and to contribute to better and more appropriate governance mechanisms in light of AI's cross sectoral nature. In light of this paper's proposal for an increase in technical capacity of AI governance institutions, diverse sets of expertise within staff members (such as backgrounds in various AI techniques, forecasting, or cybersecurity) would be advantageous to harness these new institutions' capacities. Individuals with a range of technical backgrounds could e.g.

---

[19] See for example the OECD's ONE AI network. Available here: https://www.oecd.ai/network-of-experts/.
[20] See for example the European Commission's High Level Expert Group and the working groups of the Global Partnership on AI. Available here: https://digital-strategy.ec.europa.eu/en/policies/expert-group-ai. www.https://www.gpai.ai/.

operate, manage and supervise testing or experimentation, those with a range of legal background, could e.g. work on the development of laws, those with a range of social science background could ensure governance efforts aligned with societal needs and so on, all within the same institution. This is not to say that any one background is superior to another, it is to say that *any single one background is insufficient* for the task at hand, devising good AI governance.

In order to ensure the staff has a sufficient degree of varying sets of expertise and diverse backgrounds, hiring processes and staff structures will matter a lot. While many institutions will be able to set up their own hiring processes, if you have a geographically diverse institution, it is likely that (founding) nations will expect a certain amount of representation with regards to staff within that institution (Turner 2018). These structures may already be tied to the specific nations' political agendas without considerations for diverse subject expertise and could suggest that additional options such as external networks or expert groups are required.

# Conclusion

In conclusion, this paper highlighted the importance of thinking about institution building for AI governance in more depth and provided a conceptual framework with which one can start working on this. In the axes on purpose, geography and capacity, the paper outlined both considerations and tradeoffs, and under purpose it also sketched connections to existing and future institutions. Future research directions for this topic could explore these axes for more concrete national cases or support decision making on new directions on an international level, especially as AI governance actions gain traction and clarity. There is also further scope to undertake investigation of additional axes and investigate their overlaps in more depth. Governments are under increasing pressure in light of AI development across all sectors (Whittlestone, Arulkumaran, and Crosby 2021; Schiff et al. 2020; Butcher and Beridze 2019) to come up with suitable and timely interventions and act upon them. This paper provided one proposal towards ensuring that the infrastructure we build now to action policy and other governance proposals are considered and sufficiently future proof.

**Acknowledgements:** Thanks to M. Maas, J. Whittlestone, H. Haukkala, V. Müller and the anonymous reviewers.